\documentclass[10pt,aps,prl,twocolumn,english]{revtex4-1}

\usepackage{graphicx}
\usepackage{amsmath}
\usepackage{amssymb}
\usepackage{dsfont}
\usepackage{fancyhdr}
\usepackage{yfonts}
\usepackage{braket}
\usepackage{siunitx}
\usepackage{booktabs}

\usepackage[utf8]{inputenc}
\DeclareFontFamily{U}{BOONDOX-calo}{\skewchar\font=45 }
\DeclareFontShape{U}{BOONDOX-calo}{m}{n}{
	<-> s*[1.05] BOONDOX-r-calo}{}
\DeclareFontShape{U}{BOONDOX-calo}{b}{n}{
	<-> s*[1.05] BOONDOX-b-calo}{}
\DeclareMathAlphabet{\mathcalboondox}{U}{BOONDOX-calo}{m}{n}
\SetMathAlphabet{\mathcalboondox}{bold}{U}{BOONDOX-calo}{b}{n}
\DeclareMathAlphabet{\mathbcalboondox}{U}{BOONDOX-calo}{b}{n}

\usepackage{hyperref}
\hypersetup{
	colorlinks   = true, 
	urlcolor     = blue, 
	linkcolor    = blue,
	citecolor   = blue,
}

\begin{document}

\title[]{Ground-State Cooling of a Single Atom in a High-Bandwidth Cavity}

\author{Eduardo~\surname{Uru{\~n}uela}}\email{e.urunuela@iap.uni-bonn.de}
\author{Wolfgang~\surname{Alt}}
\author{Elvira~\surname{Keiler}}
\author{Dieter~\surname{Meschede}}
\author{Deepak~\surname{Pandey}}
\author{Hannes~\surname{Pfeifer}}
\author{Tobias~\surname{Macha}}
\affiliation{Institut f{\"u}r Angewandte Physik, Universit{\"a}t Bonn, Wegelerstra{\ss}e 8, 53115~Bonn, Germany}

\begin{abstract}
We report on vibrational ground-state cooling of a single neutral atom coupled to a high-bandwidth Fabry-Pérot cavity. The cooling process relies on degenerate Raman sideband transitions driven by dipole trap beams, which confine the atoms in three dimensions. We infer a one-dimensional motional ground state population close to $90~\%$ by means of Raman spectroscopy. Moreover, lifetime measurements of a cavity-coupled atom exceeding $40$~s imply three-dimensional cooling of the atomic motion, which makes this resource-efficient technique particularly interesting for cavity experiments with limited optical access.
\end{abstract}

\maketitle
\section{I. Introduction}
Single atoms coupled to optical resonators are one of the most fundamental platforms in quantum optics and find applications in many tasks of quantum information science~\cite{Mucke2013,Nisbet-Jones2011,Chen2013,Reiserer2014,Ritter2012}. As a light-matter interface, they are a promising building block for long-distance quantum communication~\cite{Briegel1998,Duan2001} due to their ability to provide single photons of controlled shape~\cite{Nisbet-Jones2011} and to store quantum information~\cite{Koerber2018}. Ultimately, communication will always be pushed towards high rates, such that information carrying photons and corresponding resonators need to have a high bandwidth. At the same time, a strong light-matter interaction -- as required for the reversible transfer of quantum information -- has to be maintained by employing ensembles of atoms and / or decreasing the cavity mode volume~\cite{Colombe2007}. For the latter, fiber Fabry-Pérot cavities (FFPCs) 
are an attractive choice, since they also feature an intrinsic fiber coupling of the mode field~\cite{Hunger2010,Northup2014}. For optimal light-matter coupling the atom has to be confined within a fraction of the wavelength by cooling the atomic motion close to the oscillatory ground state. A standard technique in narrow-band cavities is \textit{cavity cooling}~\cite{Maunz2004,Ritsch2013}. Its steady-state temperature limit is $T_{\text{cav}}\approx \hbar\kappa/k_{\text{B}}$, where $2\kappa$ is the resonator bandwidth and $k_{\text{B}}$ the Boltzmann constant. Effective cavity cooling is therefore ineffective in the regime of high-bandwidth (i.e. open) resonators with $2\kappa$ much larger than the natural atomic linewidth $2\gamma$. In such open-cavity experiments, the optical trap depth required for trapping atoms with high equilibrium temperatures of $T_{\text{cav}}$ will be difficult to achieve.

Here, we report on an alternative cooling method based on \textit{degenerate Raman Sideband Cooling} (dRSC), which was originally developed for the loss-free cooling of neutral atom gases at high densities~\cite{Hamann1998,Vuletic1998}. We apply this method to three-dimensionally (3D) cool a single atom within the cavity mode using only dipole trap beams, a weak repumping beam and a tunable magnetic guiding field, which is a simple, resource-efficient configuration and especially beneficial for cavity experiments with limited optical access. By means of Raman spectroscopy and cavity-assisted state detection, we determine a one-dimensional (1D) ground state population close to~$90~\%$.

\section{II. Experimental Setup}
\begin{figure*}[ht]
	\centering
	\includegraphics[width=1\textwidth]{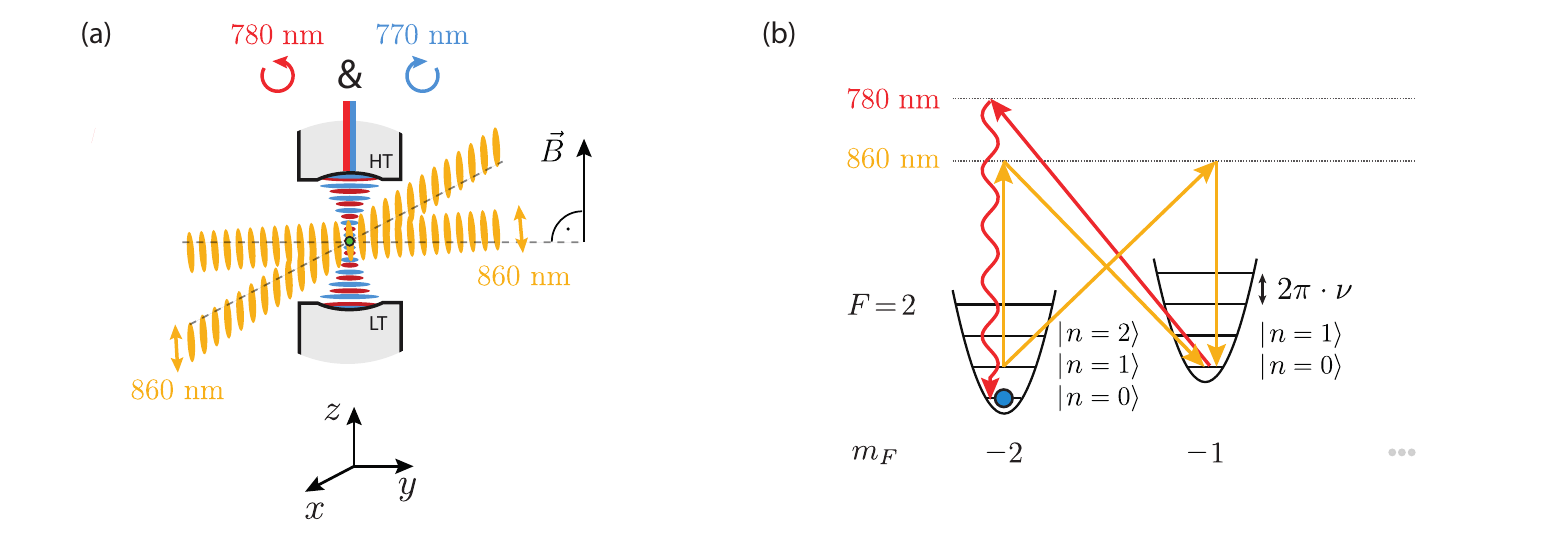}
	\caption{\textbf{(a)}~Optical fields involved in the degenerate Raman sideband cooling (dRSC) process. The $860$~nm dipole trap beams have a slightly non-orthogonal angle with respect to the cavity axis, along which the magnetic guiding field $\vec{B}$ is aligned. Two optical pumping beams enter along the dipole trap (not shown) and along the cavity axis, for the latter in the form of $780$~nm probe light entering through the high-transmission (HT) mirror. The low-transmission (LT) mirror makes for a directional, single-sided cavity. A $770$~nm lock laser is used to stabilize the resonator length via the Pound-Drever-Hall method. Its field creates an intra-cavity standing wave which traps the atoms along the $z$-direction. \textbf{(b)}~The schematic drawing shows the method of decreasing the quantized motional state $\ket{n}$ in an approximately harmonic trap potential with trap frequency $\nu$. The red-detuned dipole traps can drive $\pi$-$\sigma^-$ and $\sigma^+$-$\pi$ Raman transitions. Additionally, since the $860$~nm trap beams are not orthogonal to the cavity axis, they can couple to motional eigenstates in all directions. By optical pumping, the population in $F=1$ (not shown) and $m_F\neq-2$ states is transferred back towards $\ket{F,m_F}=\ket{2,-2}$, such that Raman cooling will be constantly active if Equation~\ref{eq:conditionfordegcooling} is fulfilled.}
	\label{setup}
\end{figure*}

Our setup consists of a single $^{87}$Rb atom trapped at the center of a high-bandwidth FFPC~\cite{Gallego2018} with CQED parameters $(g,\kappa,\gamma)=2\pi\cdot(80,41,3)$~MHz, where $g$ is the single atom-cavity coupling strength. One of the fiber mirrors has a higher transmission, providing a single-sided cavity with a highly directional input-output channel~\cite{Gallego2016}. The cavity is placed at the focus of four in-vacuum, aspheric lenses (NA = 0.5), which strongly focus two pairs of counter-propagating, red-detuned dipole trap (DT) beams at $860$~nm~\cite{Dorantes2017} in the $xy$-plane, as depicted in Fig.~\ref{setup}a. They create a 2D optical lattice, which enables atom trapping in the Lamb-Dicke regime~\cite{Leibfried2003}. One of the lattices acts as a conveyor belt~\cite{Kuhr2001} to transport single atoms from a magneto-optical trap (MOT) into the cavity. Confinement in the $z$-direction is provided by the intra-cavity, blue-detuned lock laser field at $770$~nm, which is additionally used for stabilizing the resonator length via the Pound-Drever-Hall method~\cite{Black2001}. Hence, the atom is located with sub-wavelength precision at an antinode of the cavity mode, which is driven weakly by a probe laser~\cite{Gallego2018}. The $\sigma^-$-polarized probe field and the cavity are resonant with the $\ket{F=2,m_F=-2}\rightarrow\ket{F'=3,m_F=-3}$ hyperfine transition of rubidium at $780$~nm. As a consequence, the presence of an atom is detected by an increase of the reflected probe power. A magnetic guiding field $B$ of up to $1$~G is applied along the cavity axis.

\section{III. Cooling Method}

In order to drive trap-induced, degenerate Raman transitions, the DT beams need to be able to address $\sigma^\pm$ and $\pi$-transitions simultaneously, while the Zeeman splitting $\Delta \omega_{B}$ caused by the magnetic field $B$ has to match an integer multiple $n$ of the axial trap frequency $\nu$~\cite{Grimm2000}:
\begin{equation}
\Delta \omega_{B}=n\cdot2\pi\cdot\nu\, .
\label{eq:conditionfordegcooling}
\end{equation}
In previous implementations~\cite{Vuletic1998,Hamann1998,Kerman2000}, the lattice consisted of three coplanar laser beams, two of which were linearly polarized in the lattice plane perpendicular to the quantization axis. The third one was elliptically polarized to enable Raman coupling. In our experiment, the different polarization components are generated by the geometric configuration of the dipole trap beams, see Figure~\ref{setup}{a}. The beams of DT$_{\text{x,y}}$ are slightly inclined with respect to the plane normal to the quantization axis (for DT$_{\text{x}}\leq15^\circ$ and for DT$_{\text{y}}\sim8^\circ$). Hence, the beams of the individual DTs (with linear polarization) are not purely $\pi$-polarized and $m_F$-state changing two-photon transitions are allowed.

In order to describe the Raman process, we express the internal hyperfine state $\ket{F}$ of the atom with its magnetic sublevel $\ket{m_F}$ and its excited vibrational state $\ket{n}$ as a set of discrete energy states $\Ket{F,m_F;n}$. By the combined action of the probe light and a repumper resonant with the $\ket{F=1}\rightarrow\ket{F'=2}$ transition the atom is optically pumped to the state $\ket{2,-2;n}$, see Figure~\ref{setup}{b}. 
The Raman processes are driven by DT$_{\text{x,y}}$ as $\pi$-$\sigma^-$ or $\sigma^+$-$\pi$ transitions $\ket{2,-2;n}\rightarrow\ket{2,-1;n-1}$, reducing the oscillatory quantum number $n$ by one. As a result, the atomic population is cooled into the state $\ket{2,-2;0}$, which is a dark state with respect to Raman transitions. Simultaneously, the presence of the atom is continuously interrogated by probe light. This allows to record the atom trapping lifetime $\tau$ in dependence of the Zeeman splitting $\Delta \omega_{B}$.

\section{IV. Results}

\begin{figure*}[ht]
	\centering
	\includegraphics[width=1\textwidth]{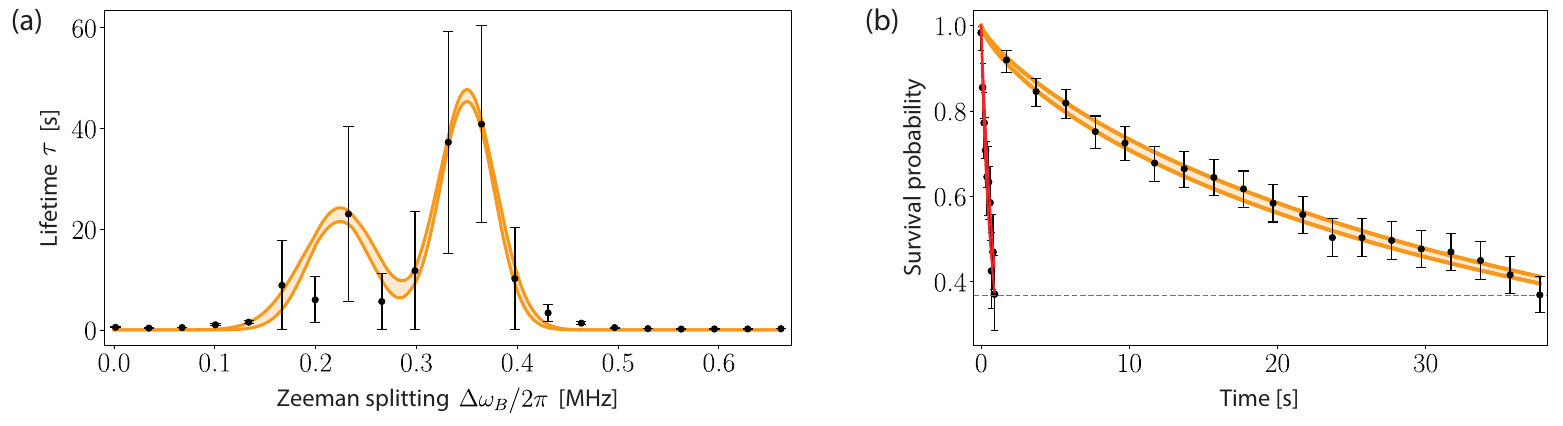}
	\caption{\textbf{(a)} A measurement of the atom trapping time $\tau$ as a function of the Zeeman splitting $\Delta \omega_{B}$. A fit of two Gaussians (yellow line) indicates dRSC whenever the absolute value of the magnetic field leads to a Zeeman level shift close to the trap frequencies. Since each measurement takes 60 seconds, only a few trials per point are available. To accurately estimate the mean lifetime, we employ the bootstrapping method, resulting in a larger margin of error.~\textbf{(b)}~Measurements of the survival probability at a fixed magnetic field lead to drastically different lifetimes depending on whether optical pumping by probe light, and thus dRSC, is present (yellow line) or absent (red line). From stretched-exponential fits (see Eq.~\ref{eq:stretchedexpo}), $1/e$ lifetimes (dashed, black line) of $(42.9\pm1.0)$~s and $(1.0\pm0.1)$~s are obtained, respectively.}
	\label{scan}
\end{figure*}

In Figure~\ref{scan}{a}, long trapping times are observed whenever the absolute value of the magnetic field leads to a Zeeman level shift on the order of the trap frequency $\nu_x,\nu_y$ or $\nu_z$, which identifies degenerate Raman transitions. From a fit of two Gaussians, the values $\nu_x=\nu_y=(350\pm1)$~kHz and $\nu_z=(224\pm5)$~kHz are extracted. The width of the Gaussians indicate inhomogeneous broadening caused by different atom positions in the 3D trapping region. Considering the optical power in the beams and the beam diameters, we estimate upper limits for the trap frequencies of $\nu_x=\nu_y=400$~kHz and $\nu_z=280$~kHz, in agreement with the measurement.

In a next step, $\Delta\omega_{B}$ is fixed to $2\pi\cdot350$~kHz, which constitutes the optimum value for cooling. Here, we investigate in more detail the survival probability for different cooling times. We find a $1/e$ lifetime of $(42.9\pm1.0)$~s by fitting the data with a stretched exponential~\cite{Williams1970,Lee2001} of the type:
\begin{equation}
e^{-(t/\tau)^k},
\label{eq:stretchedexpo}
\end{equation}
with a lifetime $\tau$ and a stretching parameter $k=(0.8\pm0.1)$. While this function is a phenomenological approach, it represents the average decay for an ensemble of decay processes with a distribution of lifetimes $\tau_i$, which depend on the inhomogeneous atom confinement in the dipole traps. In the absence of optical pumping (probe light) and thus dRSC, the average lifetime is only $(1.0\pm0.1)$~s due to heating processes induced by the cavity-resonant dipole trap. Here, a common problem is the transfer of relative frequency noise between cavity resonance and laser frequency into intra-cavity intensity fluctuations, causing additional parametric excitation of the atoms along the cavity axis~\cite{Gehm1998,Savard1997}.
 
\begin{figure}[h!]
	\centering
	\includegraphics[scale=1.02]{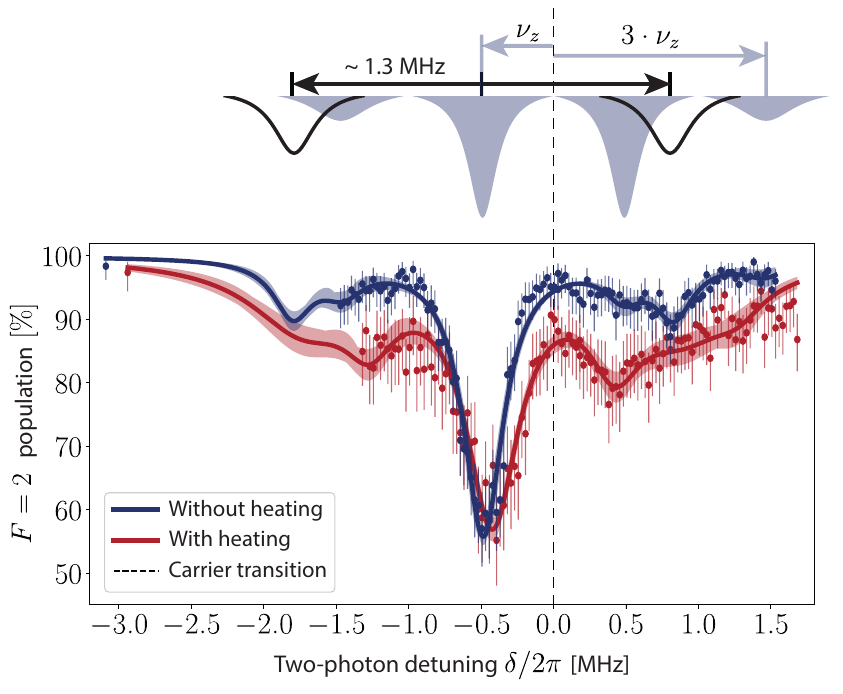}
	\caption{Carrier-free Raman spectroscopy. The atomic population not transferred by the Raman pulse, i.e. remaining in $\ket{F=2}$, is plotted versus the two-photon detuning from the hyperfine ground state splitting. A schematic drawing of the expected sidebands ($\pm\nu_z,\pm3\nu_z,...$) is shown above the measured Raman spectra. The "servo bumps" of the phase-locked loop between lock and Raman laser give rise to additional sidebands (black line in the schematic drawing above). For a Raman spectrum after dRSC (blue data points), a fit (blue line) yields a mean motional excitation along $z$ of $\overline{n}_z=(0.13\pm0.03)$, which indicates that dRSC is capable of cooling the atoms to the motional ground state. In order to elucidate the cooling sideband, we measure a second Raman spectrum (red points) after the atoms were heated by a controlled, $100$~ms long interruption of dRSC. Now, with hotter atoms, a cooling sideband becomes visible on the right side of the suppressed carrier transition (black, dashed line).}
	\label{spec}
\end{figure}

To gain insight into the temperature of the atom in the critical $z$-direction, we perform Raman spectroscopy using a second $770$~nm laser which is phase-locked to the previously introduced lock laser and enters the cavity as a running wave from the side. The Raman light has a tunable frequency offset $\delta$ around the hyperfine splitting of $+2\pi\cdot6.834$~GHz. To record a spectrum, the atom is prepared in the state $\ket{2,-2}$ by dRSC and motional state-changing transitions are driven between the states $\ket{2,-2}$ and $\Ket{1,-1}$ by a $200~\mu$s long Raman pulse, which exceeds the coherence time. During the pulse, degenerate Raman transitions are prevented by an increased magnetic field. By a cavity-assisted, non-destructive readout of the hyperfine state we measure the atomic population remaining in $\ket{F=2}$ as a function of the two-photon detuning, see Figure~\ref{spec}.
The running-wave Raman beam is sent along $y$, with DT$_y$ off, such that only odd-order sidebands along $z$ are observed. They change the motional state by $\Delta n_z=\pm1,\pm3,...$, while the carrier transition is suppressed~\cite{Reimann2014a,Neuzner2018}. The noise peaks ("servo bumps") of our Raman laser phase-locked loop appear as additional features at $\pm 1.3$~MHz of any Raman transition, but they are mainly visible for the strong heating sideband. The depths of the dips depend on the technical details of the Rabi spectroscopy pulse and do not play a role in calculating the mean motional excitation number $\overline{n}_i$ along the direction $i$. Assuming a thermal equilibrium, $\overline{n}_i$ is given by the relation
\begin{equation*}
\overline{n}_i=\frac{R_i}{1-R_i}\, ,
\end{equation*}
where $R_i$ is the ratio of the areas under the cooling and the heating sideband. Since in the presented spectra the sidebands overlap, we extract this ratio from a fit considering all expected sidebands. The resulting mean motional excitation along $z$ is $\overline{n}_z=(0.13\pm0.03)$. This corresponds to a one-dimensional ground state population $n_{0,z}=1/(1+\overline{n}_z)$ of $(88\pm3)~\%$. To validate our interpretation of the Raman spectrum, we record a second spectrum with atoms at higher temperatures by introducing a $100$~ms long waiting time before each spectroscopy pulse during which the atoms heat up. As a consequence, the cooling sideband becomes clearly visible. In this case, the mean motional quantum number is $\overline{n}_z=(0.47\pm0.06)$.

\section{V. Conclusion}

We have applied a simple and robust method to cool a single atom inside a high-bandwidth resonator to its one-dimensional motional ground-state. The long trapping lifetime of $40$~s under continuous, non-destructive probing of the atom's presence allows interesting applications such as determining the atomic position within the cavity by imaging the probe light scattered into free space. 
Without cooling, the atom trapping lifetime in each 1D red dipole trap is limited by phase-noise to only $\sim15$~s. Thus, the observation of a significantly longer lifetime suggests that the atoms are cooled in three dimensions.

Since only weak optical pumping and a tunable magnetic bias field are required, the dRSC method has the potential to complement established techniques such as cavity cooling -- even for narrow-band cavities where cavity cooling works well. 
It is worthwhile to point out that the tools used for Raman spectroscopy can directly be applied for carrier-free ground-state Raman cooling in three dimensions, if the Raman beam is sent diagonally in the $xy$-plane. In our setup, this method will supersede dRSC as soon as the cooling conditions need to be (de)activated faster than the timescale on which the magnetic field can be changed.

\section{Acknowledgments}

This work has been funded by the Deutsche Forschungsgemeinschaft (DFG, German Research Foundation) - Project number 277625399- TRR 185 and the Bundesministerium für Bildung und Forschung (BMBF), Verbundprojekt Q.Link.X.

\bibliography{Biblio}
\bibliographystyle{apsrev4-1}
\end{document}